\documentclass[nofootinbib,showpacs,aps,twocolumn,preprintnumbers,letterpaper]{revtex4-2}
\usepackage{amsmath,amssymb}
\usepackage{epsfig}
\usepackage{graphicx}
\usepackage{amsmath}
\usepackage{slashed}
\usepackage{amsfonts}
\usepackage{epstopdf}
\usepackage{tablefootnote}
\usepackage[normalem]{ulem}

\usepackage{graphicx,subfigure}
\usepackage{amsmath,amssymb}
\usepackage{hhline}

\usepackage[pdftex, pdfborder= 0 0 0, citecolor=blue, urlcolor=blue, linkcolor=blue, colorlinks=true, bookmarksopen=true]{hyperref}

\newcommand{\be}{\begin{eqnarray}}
	\newcommand{\ee}{\end{eqnarray}}

\def\slashchar#1{\setbox0=\hbox{$#1$}           
	\dimen0=\wd0                                 
	\setbox1=\hbox{/} \dimen1=\wd1               
	\ifdim\dimen0>\dimen1                        
	\rlap{\hbox to \dimen0{\hfil/\hfil}}      
	#1                                        
	\else                                        
	\rlap{\hbox to \dimen1{\hfil$#1$\hfil}}   
	/                                         
	\fi}                                         %

\begin{document}
	
	\title{Low-energy enhancement of the magnetic dipole radiation in odd-mass lanthanides}
	
	\author{ D. DeMartini  and Y. Alhassid}
	
	\affiliation{Center for Theoretical Physics, Sloane Physics Laboratory, Yale University, New Haven, Connecticut 06520, USA}
	
\begin{abstract}
We compute the magnetic dipole (M1) $\gamma$-ray strength functions ($\gamma$SF)  for the odd-mass lanthanides $^{\textrm{143-151}}$Nd and $^{\textrm{147-153}}$Sm using the shell-model Monte Carlo method in combination with the static-path approximation and the maximum-entropy method.  In particular, we quantify the statistical uncertainties in the calculated M1 $\gamma$SFs and show that they are under control for the excitation energies relevant to the experiments despite a Monte Carlo sign problem that originates in the projection onto an odd number of neutrons. We identify a low-energy enhancement (LEE) in the M1 $\gamma$SFs of these odd-mass lanthanides, which was recently observed experimentally in some of them. We also find a scissors mode resonance (SR) in the strongly deformed isotopes.  We observe that the decrease in the LEE strength with neutron number along an isotopic chain is compensated for by an increase in the SR strength in the deformed nuclei. We compare our results with recent experiments.
\end{abstract}

\maketitle
	
\section{Introduction}
	
The absorption and emission spectra of $\gamma$-rays from atomic nuclei reveal important information about collective structures in these nuclei. The average  reduced $\gamma$-decay probability, known as the $\gamma$-ray strength function ($\gamma$SF\footnote{The $\gamma$SF is also referred to as the photon strength function (PSF) or the radiative strength function (RSF).}), plays a crucial role in compound-nucleus reactions~\cite{Hilaire:2021djf} such as neutron capture in stellar nucleosynthesis~\cite{Arnould:2007gh}. 
	
 The most dominant feature of the $\gamma$SFs is the giant dipole resonance (GDR), an electric dipole (E1) resonance.  However, other smaller structures have been observed at lower energies. The scissors resonance (SR), which describes the oscillations of the neutron and proton clouds with respect to each other like scissors blades, was observed in the magnetic dipole (M1) strength of deformed nuclei at $E_{\gamma} \sim 2- 3$ MeV. At higher excitation energies $E_{\gamma} \sim 6-9$ MeV, the M1 $\gamma$SF is dominated by the spin-flip mode~\cite{RevModPhys.82.2365}.

A previously unexpected feature in the $\gamma$SF, an `upbend' in the strength in the lowest $\gamma$-ray energies, was discovered in several light- and medium-mass nuclei in the past two decades~\cite{Voinov:2004pb,Guttormsen:2004xj,Wiedeking:2012zz,Larsen:2013xks,Larsen2018}. This feature, now known as the low-energy enhancement (LEE), was confirmed to be dipole in nature via measurement of angular correlations~\cite{Larsen:2013xks,Jones:2018zta}, but it has not yet been determined experimentally whether it is an M1 or an E1. Conventional configuration-interaction (CI) shell-model calculations identified a LEE in the M1 $\gamma$SF~\cite{Schwengner:2013ora,Brown:2014bta,Schwengner:2017cet,Sieja2017,Karampagia:2017zgj,Sieja2018,Frauendorf:2021gfd}, although it was shown in Ref.~\cite{Litvinova:2013oda} that nucleonic transitions from the thermally unblocked quasi-particle states to the quasi-particle continuum can lead to an enhancement in the $E1$ strength function at low $\gamma$-ray energies.

There are several experimental techniques for measuring $\gamma$SFs; see Ref.~\cite{Goriely:2019kzz} for an overview of various methods. Many of the experiments that observed a LEE used the Oslo method~\cite{SCHILLER2000498}, which determines simultaneously the nuclear level density and the $\gamma$SF below the neutron separation energy.

Recently,  Oslo-method experiments have been extended to nuclei in the rare-earth region. The LEE was observed in the odd-mass samarium isotopes $^{\textrm{147-153}}$Sm~\cite{Simon:2016mif,Naqvi:2019odv}\footnote{A subsequent experiment found no evidence of the LEE in $^{153}$Sm~\cite{Malatji:2021nar}.} and in even- and odd-mass neodymium isotopes $^{\textrm{142,144-148}}$Nd~\cite{Guttormsen:2022ccz}. If the LEE persists in heavy, neutron-rich nuclei near the drip line, it would enhance the $(n,\gamma)$ cross sections in $r$-process nucleosynthesis by more than an order of magnitude~\cite{Larsen:2010ayz}.

 Conventional CI shell model calculations become intractable in heavy open-shell nuclei due to the combinatorial increase in the size of the many-particle space with the number of valence nucleons and/or the number of single-particle states used in the calculation. The shell-model Monte Carlo (SMMC) method~\cite{Johnson1992,Lang1993,Alhassid1994} has been used successfully to calculate finite-temperature observables in lanthanides~\cite{Alhassid2008,Ozen:2012gq,Ozen:2013xza}. However, the $\gamma$SF cannot be calculated directly in SMMC. It is only possible to compute its Laplace transform, the imaginary-time response function. Calculation of the strength function from the imaginary-time response function requires a numerical analytic continuation. This analytic continuation can be carried out in the maximum-entropy method (MEM)~\cite{Jarrell:1996rrw}.  
 
A successful application of the MEM requires a good prior strength function. In Ref.~\cite{Fanto2024} the static-path plus random-phase approximation (SPA+RPA)~\cite{PhysRevC.42.R1830,PUDDU1991409,ATTIAS1997565,PhysRevLett.80.1853} was used as a prior strength function for a pairing plus quadrupole interaction to calculate M1 $\gamma$SFs in even-mass samarium isotopes. However, for more general interactions, the calculation of the SPA+RPA strength function becomes too time consuming, and in Ref.~\cite{Mercenne2024} the SPA strength function was used as a prior to calculate the M1 $\gamma$SFs in an isotopic chain of even-mass neodymium nuclei.  The SPA~\cite{Muhlschlegel1972, Alhassid1984, Lauritzen1988,Arve1988,Rossignoli:1992lpw,Alhassid:1992uyi,ATTIAS1997565} is a simplification of the SPA+RPA that only takes into account large-amplitude static fluctuations of the fields and ignores the time-dependent fluctuations. In the SPA, unlike in the SMMC, the strength function can be calculated directly. 

Here, we apply this combination of the SMMC and SPA methods to calculate the M1 $\gamma$SFs in isotopic chains of heavy odd-mass neodymium nuclei $^{\textrm{143-151}}$Nd and samarium nuclei $^{\textrm{147-153}}$Sm. The projection onto an odd number of neutrons leads to a Monte Carlo sign problem,  which can result in large fluctuations of the response functions and make the calculations unreliable. It is therefore necessary to quantify the uncertainties in the $\gamma$SFs when carrying out the analytic continuation using the MEM. 

In the M1 $\gamma$SFs near the neutron separation energy, we clearly identify a LEE in all of the odd-mass samarium and neodymium isotopes under study.  We also observe the SR in the deformed isotopes and a spin-flip mode.

 We compare our results with recent Oslo-method experiments that extracted the $\gamma$SFs in these isotopes.  For the neodymium isotopes in which a LEE was observed experimentally, we find overall agreement with our calculations. In contrast, for the samarium isotopes the strength of the LEE at $E_\gamma\approx 0$  is significantly larger than its calculated value. We note though that our theoretical results for the LEE in the samarium isotopes are similar in magnitude to those for the neodymium isotopes.
	
The outline of this article is as follows: in Sec.~\ref{sec_gamma}, we discuss the M1 deexciation $\gamma$SF (denoted by $f_{M1}$) and relate it to the M1 strength function $S_{M1}$ which is more suitable for theoretical calculations.  In Sec.~\ref{sec_SMMC}, we briefly present the SMMC method and the choice of model space and interaction parameters. In Sec.~\ref{sec_SPA}, the SPA and MEM and their roles in computing the strength functions are discussed. In Sec.~\ref{sec_results} we present the main results of this work. First we introduce the method for computing the uncertainties of the strength functions and show how those uncertainties vary with temperature. Then, we present results for the M1 strength functions $S_{M1}$ in odd-mass samarium and neodymium isotopes for temperatures near the ground state and at the neutron separation energy. The latter are converted to the deexcitation $\gamma$SF $f_{M1}$ and  compared with the recent Oslo-method experiments. Finally, in Sec.~\ref{sec_sum} we summarize the results and discuss future potential improvements of the calculations. 

\section{$\gamma$-ray Strength Functions} \label{sec_gamma}
	
In this section we discuss $\gamma$-ray strength functions. In particular we focus our discussion on M1 transitions.

\subsubsection{M1 strength functions} 

In experiments, one usually measures the $\gamma$-ray decay of the compound nucleus from an initial energy $E_i$ and spin-parity $J_i^\pi$.  The deexcitation M1 $\gamma$SF  $f_{M1}$ is defined by the relation 
\be\label{def_f_M1}
 {\langle \Gamma_{M1}(E_i,J^{\pi}_i; E_\gamma) \rangle \over  D_{J_i^\pi}} = E_\gamma ^3  \, f_{M1}(E_i,J^{\pi}_i; E_\gamma)\;,
\ee
where $\langle \Gamma_{M1}(E_{\gamma};E_i,J^{\pi}_i) \rangle $ is the average partial M1 $\gamma$-ray decay width for a compound nucleus at an initial excitation energy $E_i$ and spin-parity $J^{\pi}_i$ to emit a photon with energy $E_{\gamma}$ and $D_{J_i^\pi}$ is the average level spacing of the initial states~\cite{Bartholomew1973}. 

Theoretically, it is advantageous to define the M1 strength function $S_{M1}$ at temperature $T$  and initial spin-parity $J^\pi_i$  by
\begin{equation}\label{S_M1}
	\begin{aligned}
		S_{M1}(E_i,J^{\pi}_i, \omega) &= \sum_{\alpha_i} \sum_{\alpha_f  J_f} \sum_{M_i,M_f,\mu} \frac{1}{2 J_i + 1} \frac{e^{-\beta E_{\alpha_i J_i}}}{Z} \\
		&\times |\langle f | \hat{\mathcal{O}}_{M1}^{\mu} | i \rangle | ^2 \delta (\omega + E_i - E_f) \;,
	\end{aligned}
\end{equation}
where $\omega$ is the transition energy and $i\equiv \alpha_i\; J_i \;M_i $ denote eigenstates of the CI shell-model Hamiltonian $\hat{H}$ with energies $E_i$, total spins $J_i$, and total magnetic quantum numbers $M_i$. Here, $\hat{\mathcal{O}}_{M1}^{\mu} $ is the $\mu$ component of the M1 transition operator, and $Z=\sum_{\alpha_i J_i} (2J_i+1) e^{-\beta E_{\alpha_i J_i}}$ is the canonical partition function at an inverse temperature $\beta=1/T$. This expression can be simplified by summing explicitly over the magnetic quantum numbers to find \begin{eqnarray}\label{eq_strength}
\!\!\!\! S_{M1}(T; \omega) = \sum_{\substack{\alpha_i \\ \alpha_f J_f}} &\frac{e^{-\beta E_{\alpha_i J_i}}}{Z}   |(f || \hat{\mathcal{O}}_{M1} || i)|^2 \nonumber \\  & \times \delta(\omega - E_f + E_i) \;,
\end{eqnarray}
where $(f || \hat{\mathcal{O}}_{M1} || i)=(\alpha_f J_f || \hat{\mathcal{O}}_{M1} || \alpha_i J_i )$ are the reduced matrix elements. 
The bare M1 operator is given by
\begin{equation}
	\hat{\mathcal{O}}_{M1} = \sqrt{\frac{3}{4 \pi}} \frac{\mu_N}{\hbar c} (g_l \hat{\bf l} + g_s \hat{\bf s}),
\end{equation}
where $\hat{\bf l}$ and $\hat{\bf s}$ are, respectively, the orbital and spin angular momentum operators. The free-nucleon $g$-factors used here are $g_{l,p} = 1$, $g_{l,n} = 0$, $g_{s,p} = 5.5857$, and $g_{s,n} = -3.8263$. The reduced matrix elements in Eq.~(\ref{eq_strength}) are nonzero only for those with final states satisfying the M1 selection rules: $J_f = J_i$, $J_i \pm1$ and $\pi_f = \pi_i$.

In the following, we relate the deexcitation $\gamma$-ray strength function $f_{M1}$ to the strength function $S_{M1}$.  The partial width $\Gamma_{M1}(E_i,J^{\pi}_i; E_\gamma)$ for an M1 $\gamma$-ray decay is given by 
\be
\Gamma_{M1}(E_i,J^{\pi}_i; E_\gamma) = a E_{\gamma}^3 | \langle f || \hat{\mathcal{O}}_{M1} || i \rangle |^2  \;,
\ee
where $a=\frac{16 \pi}{9 (\hbar c)^3}$. Using this relation in Eq.~(\ref{def_f_M1}), we obtain the following expression for the deexcitation M1 strength function
\be
 f_{M1}(E_{\gamma};E_i,J^{\pi}_i) = a \tilde{\rho}(E_i,J^{\pi}_i) \langle |(f || \hat{\mathcal{O}}_{M1} || i )|^2 \rangle \;,
 \ee
 where $ \tilde{\rho}(E_i,J^{\pi}_i)$ is the average level density at energy $E_i$ and spin-parity $J^{\pi}_i$, and $\langle \ldots\rangle$ denotes an average over both initial and final states. Integrating (\ref{S_M1}) over a narrow energy window $\Delta \omega$, we can relate the average $\langle |(f || \hat{\mathcal{O}}_{M1} || i )|^2 \rangle$ to a smoothed strength function $\bar S_{M1}$
 \be
 \langle |(f || \hat{\mathcal{O}}_{M1} || i )|^2 \rangle \approx \frac{1}{\tilde \rho_f } \bar S_{M1}(T; \omega = -E_\gamma) \;,
 \ee
 where we have replaced the uniform average over initial states by a thermal average at temperature $T$ that corresponds to an average energy $E_i$ and  $\tilde \rho_f$ is the average final level density. Using the M1 selection rules, we find for $J_i\neq 0$ (and suppressing the parity) that $\tilde \rho_f = \tilde{\rho}(E_f,J_i-1) + \tilde{\rho}(E_f,J_i) + \tilde{\rho}(E_f,J_i+1)$. We conclude
 \begin{eqnarray}
f_{M1}(E_i,J^{\pi}_i; E_{\gamma}) = & a  \frac{\tilde{\rho}(E_i,J^{\pi}_i)}{\sum_{J_f} \tilde \rho(E_f=E_i-E_\gamma,J_f) } \nonumber \\ & \times \bar S_{M1}(T,J_i^\pi; \omega = -E_\gamma) \;,
 \end{eqnarray}
 where the sum is over the three possible final spin values for any given initial spin $J_i \neq 0$.  We note that negative values of the transition energy $\omega$ in $S_{M1}$   correspond to the emission of a photon (i.e., $\gamma$-ray decay) while positive values correspond to the absorption of a photon.

In experiments such as those using the Oslo method, the compound nucleus is formed in nuclear reactions that lead to a range of initial spins and energies. Typically these experiments average over all initial spins and an initial energy window $E_{min} < E_i < S_n$, where $E_{min}$ is a lower bound $\sim 2-3$ MeV and $S_n$ is the neutron separation energy. This reduces the deexcitation $\gamma$SF to a function depending only on $E_{\gamma}$. 

The generalized Brink-Axel (gBA) hypothesis~\cite{Brink,Axel} states that the $\gamma$SF is  solely a function of the $\gamma$-ray energy and is independent of the initial energy, spin, and parity of the nucleus.  The gBA hypothesis was originally proposed in the context of the E1 giant dipole resonance but has since been applied to other electromagnetic transitions.   Recent experiments have claimed its validity in different nuclei over a variety of energy ranges~\cite{PhysRevLett.116.012502,CrespoCampo:2018bde}, while measurements of the $^{90}$Zr $\gamma$SF have found deviations from the gBA hypothesis~\cite{Netterdon:2015qca}. Theoretical results suggest that  in general the strength functions do depend on the initial energy~\cite{Johnson:2015era}.
 
 Techniques to calculate the spin dependence of the M1 $\gamma$SF are yet to be developed within the SMMC method. We therefore assume here for simplicity that the M1 strength function is independent of the initial spin-parity $J_i^{\pi}$. The ratio between the corresponding initial and final level densities can be estimated using the spin cutoff model. We find that this ratio is well approximated by $\frac{1}{3} \rho(E_i)/\rho(E_f)$, where the factor of $1/3$ accounts for the decay from a given initial spin value $J_i$ into three possible final spin values $J_f=J_i, J_i\pm 1$. The final expression for the deexcitation M1 strength function  is  then
 \begin{equation}
 	\label{eq_gammaSF}
 	f_{M1}(E_i; E_{\gamma}) \approx  \frac{1}{3} a  \frac{\tilde{\rho}(E_i)}{\tilde \rho(E_i-E_\gamma)} \bar S_{M1}(T; \omega = -E_\gamma) \;.
 \end{equation}

In Eq.~(\ref{eq_gammaSF}) we do not assume independence of the M1 $\gamma$SF on the initial energy $E_i$. In Sec.~\ref{sec_results} we will show that above a certain low value of the initial energy, our SMMC results for the M1 $\gamma$SF support the independence on initial energy of the gBA hypothesis.  We note that instead of working in the micro-canonical ensemble  of fixed initial energy $E_i$,  we use the canonical ensemble with a fixed temperature $T$ that is determined  by the condition that average excitation energy is the given value of $E_i$, i.e.,  $\langle E_x \rangle = E_i$. 

\subsubsection{Imaginary-time response function}

An important quantity which is related to the M1 strength function $S_{M1}$, is the M1 response function $R_{M1}$ (also called the imaginary-time correlation function). This response function characterizes the correlation between the value of the M1 operator at an arbitrary Euclidean time $\tau$ and at $\tau=0$ and is defined by
\begin{equation}
	R_{M1}(T; \tau) = \langle \hat{\mathcal{O}}_{M1}(\tau) \hat{\mathcal{O}}_{M1}(0) \rangle \;,
\end{equation}
where $\hat{\mathcal{O}}_{M1}(\tau) = e^{\tau H}\hat{\mathcal{O}}_{M1} e^{-\tau H}$. The M1 response function is the Laplace transform of the M1 strength function. Using the symmetry relation $S_{M1}(T; -\omega) = e^{-\beta \omega}S_{M1}(T;\omega)$,  we can rewrite
\begin{equation}
	\label{eq_response}
	R_{M1}(T;\tau) = \int_0^{\infty} d\omega K(\tau,\omega) S_{M1}(T; \omega),
\end{equation}
where $K(\tau,\omega) = e^{-\tau \omega} + e^{(\beta - \tau) \omega}$ is a symmetrized kernel. The total integrated strength is just the response function at $\tau = 0$
\begin{equation}
\int_{-\infty}^{\infty} d\omega S_{M1}(T;\omega) = \langle (\hat{\mathcal{O}}_{M1}(0))^2 \rangle = R_{M1}(T;0) \;,
\end{equation}
and moments of the strength function can be related to derivatives of the response function at $\tau = 0$. The advantage of the M1 response function is that it  can be calculated exactly within the SMMC framework, while the M1 strength cannot.

Because the kernel in Eq.~(\ref{eq_response}) decays exponentially in $\tau$, it is straightforward to calculate $R_{M1}$ given $S_{M1}$. However, the inverse problem of determining numerically $S_{M1}$ given $R_{M1}$ is ill-poised given the insensitivity to large values of $\omega$. For a given set of numerical response function data, there are  an infinite number of strengths that can satisfy the integral relation (\ref{eq_response}) within the uncertainties. 

The inversion of Euclidean response function data in order to determine strength or spectral functions is a ubiquitous problem in many-body quantum mechanics and lattice quantum field theory. Many different techniques have been proposed to solve this problem, all of which have had varying levels of success depending on the physical system to which they are applied. See Ref. \cite{Tripolt:2018xeo} and references within for discussions and comparisons of some of the techniques in the context of lattice field theory.
	
\section{Shell-model Monte Carlo} \label{sec_SMMC}
	
\subsection{The auxiliary-field Monte Carlo method}
	
The microscopic description of nucleons occupying nuclear shells with residual two-body interactions, known as the CI shell model approach, has been used to account for strong correlations in nuclei for decades. However, for heavy open-shell nuclei, the dimensionality of the many-particle CI model space is prohibitively large for conventional diagonalization methods to work, even on modern supercomputers. Such large model spaces can be studied in the auxiliary-field Monte Carlo (AFMC) approach~\cite{Alhassid2017}, known in nuclear physics as the SMMC method. 
	
The CI shell model Hamiltonian $\hat{H}$ is composed of a one-body part describing a set of single-particle orbitals with energies $\epsilon_i$ and residual two-body interactions.  It can be written in the diagonal form
\begin{equation}
	\hat{H} = \sum_i \epsilon_i \hat{n}_i + \frac{1}{2} \sum_{\alpha} v_{\alpha} \hat{\rho}_{\alpha}^2,
\end{equation}
where $\hat{n}_i$ is the one-body occupation in state $i$, $\hat{\rho}_{\alpha}$ are linear combinations of the one-body densities, and $v_{\alpha}$ are the interaction eigenvalues.
The Gibbs operator $e^{-\beta \hat{H}}$ describes the many-body evolution operator of the system at an inverse temperature $\beta = 1/T$. The crux of the AFMC is the Hubbard-Stratonovich (HS) transformation~\cite{1957SPhD....2..416S,Hubbard:1959ub}, which represents this propagator as an integral of one-body propagators in the presence of imaginary-time-dependent fields $\sigma_{\alpha}(\tau)$.
\begin{equation}
	e^{-\beta \hat{H}} = \int D[\sigma]G_{\sigma}\hat{U}_{\sigma} \;,
\end{equation}
where $G_{\sigma}$ is a Gaussian weight, $\hat{U}_{\sigma}$ is the one-body propagator describing non-interacting nucleons in time-dependent fields, and $D[\sigma]$ is a measure. The thermal expectation value of an observable $\hat{O}$ can then be written in the form
 \begin{equation}
    \langle \hat{ O } \rangle = \frac{ \int_{} D[\sigma] { W }_{ \sigma } { \Phi }_{ \sigma } { \langle O \rangle }_{ \sigma } }{ \int_{} D[\sigma] { W }_{ \sigma } { \Phi }_{ \sigma } } \;,
    \label{eq_obs}
  \end{equation}
  where ${ { W }_{ \sigma } = { G }_{ \sigma } |\text{Tr} { \hat{ U } }_{ \sigma }| }$ is a positive-definite weight, ${ { \Phi }_{ \sigma } = \text{Tr\,} { \hat{ U } }_{ \sigma } / |\text{Tr\,} { \hat{ U } }_{ \sigma }| }$ is the Monte Carlo sign function, and ${ { \langle O \rangle }_{ \sigma } = \text{Tr}(\hat{ O } { \hat{ U } }_{ \sigma })/ \text{Tr} { \hat{ U } }_{ \sigma } }$.  The one-body propagator $\hat{U}_{\sigma}$ can be represented in the single-particle space by an $N_s \times N_s$ matrix $\boldsymbol{U_{\sigma}}$, where $N_s$ is the number of single-particle states in the model space. Its many-particle trace in Fock space can be expressed in terms of $\boldsymbol{U_{\sigma}}$
  \begin{equation}
	\textrm{Tr} \hat{U}_{\sigma} = \det (\boldsymbol{1} + \boldsymbol{U_{\sigma}}) \;.
\end{equation}
This trace is essentially the grand-canonical partition function of non-interacting fermions in time-dependent auxiliary fields $\sigma(\tau)$.  For a nucleus with given numbers of valence protons $N_p$ and neutrons $N_n$, the relevant ensemble is the canonical ensemble. Canonical expectation values of observables are computed by an exact particle-number projection via a discrete Fourier transform. For a given number of valence fermions $A$, the canonical partition function is~\cite{Ormand:1993as}
\begin{equation}
	\textrm{Tr}_{A}\hat{U}_{\sigma} = \frac{e^{-\beta \mu A}}{N_s} \sum_{m=1}^{N_s} e^{-i \varphi_m A} \det (\boldsymbol{1} +e^{i \varphi_m + \beta \mu} \boldsymbol{U_{\sigma}}),
\end{equation}  
where $\varphi_m = 2 \pi m / N_s$ $(m= 1,...,N_s)$ are quadrature points and $\mu$ is a chemical potential used to stabilize numerically the Fourier sum. The chemical potential $\mu$  is chosen to reproduce the correct average number of valence particles. In the calculations we use two projections, one for protons and another for neutrons. 

Monte-Carlo sampling is carried out using the Metropolis algorithm in the space of auxiliary-field configurations $\sigma_{\alpha}(\tau_n)$. The samples are chosen according to the positive-definite weight function $W_{\sigma} = G_{\sigma} | \textrm{Tr}_A \hat{U}_{\sigma}|$.  Quantum Monte Carlo for fermions often  have a sign problem when the sign $\Phi_{\sigma} = \textrm{Tr}_A \hat{U}_{\sigma}/|\textrm{Tr}_A \hat{U}_{\sigma}|$  is negative for a significant fraction of the samples, in which case the statistical fluctuations in the calculated observables can become too large. However, there are good-sign interactions for which the grand-canonical trace is positive for all auxiliary-field configurations $\sigma$. These interactions include the dominant collective attractive components of effective nuclear interactions~\cite{Dufour1996} and have been used successfully to calculate statistical properties of nuclei.  The smaller bad-sign components of effective nuclear interactions can be treated using the extrapolation method introduced in Ref.~\cite{Alhassid1994}.
While the projection on even number of particles keeps the sign good, the projection on an odd number of particles leads to a sign problem at low temperatures, making it impossible to calculate directly the ground-state energy  of a nucleus with odd numbers of protons and/or neutrons. Methods were developed to extract the ground-state energy of such nuclei despite the odd particle-number sign problem~\cite{Ozen:2013xza,Alhassid2023}. 

The imaginary-time M1 response function is calculated using 
\begin{equation}\label{response-SMMC}
  { R }_{ M1} (T;\tau) = \frac{ \int_{} D[\sigma] W_\sigma \Phi_\sigma  { \langle \hat{\mathcal{O}}_{M1}(\tau) \cdot \hat{\mathcal{O}}_{M1} \rangle }_{ \sigma } }{ \int D[\sigma] W_\sigma\Phi_\sigma} \;,
\end{equation}
  where $ \hat{\mathcal{O}}_{M1}(\tau) \cdot \hat{\mathcal{O}}_{M1}=\sum_\mu (-)^\mu \hat{\mathcal{O}}^\mu_{M1}(\tau)\hat{\mathcal{O}}^{-\mu}_{M1}$ and ${\hat{\mathcal{O}}_{M1} (\tau) = { \hat{ U } }_{ \sigma }^{ -1 } (\tau,0) \hat{\mathcal{O}}_{M1}{ \hat{ U } }_{ \sigma } (\tau,0) }$ with $\hat U_\sigma(\tau,0)$ being the propagator between times $0$ and $\tau$ for a given sample $\sigma$. The expectation value ${ { \langle \dots \rangle }_{ \sigma } }$ is taken with respect to the propagator $\hat U_\sigma$ between $\tau=0$ and $\tau=\beta$.
  
\subsection{Model space and Hamiltonian}
	
For the rare-earth nuclei studied in this work, we use the same model space and Hamiltonian as in previous studies of even- and odd-mass isotopes in this mass region~\cite{Ozen:2012gq,Ozen:2013xza,Guttormsen:2020nnp}. The single-particle model space includes the $0g_{7/2}$, $1d_{5/2}$, $1d_{3/2}$, $2s_{1/2}$, $0h_{11/2}$, and $1f_{7/2}$ orbitals for protons, and the $0h_{11/2}$, $0h_{9/2}$, $1f_{7/2}$, $1f_{5/2}$, $2p_{3/2}$, $2p_{1/2}$, $i_{13/2}$, and $1g_{9/2}$ orbitals for neutrons. Including magnetic degeneracy, there are a total of 40 proton single-particle states and 66 neutron states. The single-particle energies are obtained from the spherical Woods-Saxon potential plus spin-orbit coupling. Effects of the inert $^{120}$Sn core are included via renormalization of the interaction terms. 
	
The effective residual interactions consist of monopole pairing and multipole terms
\begin{equation}
		\hat{H}_{int} = -\sum_{\nu} g_{\nu} \hat{P}_{\nu}^{\dagger} \hat{P}_{\nu} - \sum_{\lambda} \chi_{\lambda} (\hat{\mathcal{O}}_{\lambda,p}+\hat{\mathcal{O}}_{\lambda,n}) \cdot (\hat{\mathcal{O}}_{\lambda,p}+\hat{\mathcal{O}}_{\lambda,n}),
\end{equation}
where $\nu = p,n$ is the nucleon species and $\lambda=2,3,4$ corresponds to the $2^{\lambda}$-pole interactions. The pair creation operator $\hat{P}_{\nu}^{\dagger}$ and multipole operator $O_{\lambda,\nu}$ are given by
\begin{equation}
	\hat{P}_{\nu}^{\dagger} = \sum_{n\ell Jm} (-1)^{J+m+\ell} a_{n \ell J m, \nu}^{\dagger} a_{n \ell J -m, \nu}^{\dagger},
\end{equation}
\begin{equation}
	\hat{\mathcal{O}}_{\lambda,\nu} = \frac{1}{\sqrt{2 \lambda + 1}} \sum_{ab} \langle J_a | \frac{dV_{WS}}{dr} Y_{\lambda} | J_b \rangle [a_{n \ell J_a , \nu}^{\dagger} \times \tilde{a}_{n \ell J_b , \nu}^{\dagger}]^{\lambda},
\end{equation}
where $\tilde{a}_{Jm} = (-1)^{J+m} a_{J-m}$ and $V_{WS}$ is the central Woods-Saxon potential. The pairing strength $g_{\lambda} = \gamma \bar{g}_{\lambda}$ with $\bar{g}_p = 10.9/Z$, $\bar{g}_n = 10.9/N$. The interaction strengths are $\chi_{\lambda} = k_{\lambda} \chi$. The parameter $\chi$ is determined self-consistently from the invariance of the one-body potential under a displacement of the nucleus \cite{Alhassid:1995tt}, while $k_{\lambda}$ are renormalization factors that take into account the effects of the core. The remaining parameters are $\gamma = 0.72 - 0.5/((N-90)^2 + 5.3)$, $k_2 = 2.15 + 0.0025(N-87)^2$, and $k_3 = k_4 = 1$. 

\subsection{Computation of observables in SMMC}

Thermal averages of observables (\ref{eq_obs}) are computed by averaging over sets of thermalized and uncorrelated field configurations $\sigma_i$ chosen by the Metropolis algorithm according to the weight function $W_\sigma$ with Monte-Carlo signs $\Phi_{\sigma_i}$
\begin{equation}
	\langle \hat{O} \rangle = \frac{\sum_i \langle \hat{O} \rangle_i \Phi_{\sigma_i}}{\sum_i \Phi_{\sigma_i}}\;.
\end{equation} 

The SMMC calculations of observables are done by taking discrete Euclidean time slices of width $\Delta \beta$. In order to eliminate systematic error due to this discretization, we perform calculations at two values of $\Delta \beta = 1/32$ MeV$^{-1}$ and $1/64$ MeV$^{-1}$ and thermal averages are linearly extrapolated to $\Delta \beta \rightarrow 0$ at a fixed value of $\beta$. For the thermal energy $E(\beta)$,   the dependence on $\Delta \beta$ is weaker at larger values of $\beta \geq 3$ MeV$^{-1}$, and an average is used instead of extrapolation to reduce the statistical uncertainty. For the M1 response function $R_{M1}$, a moderate dependence on $\Delta \beta$ remains at large $\beta$, and a linear extrapolation is used for all temperatures.  

\section{Techniques for computing the strength functions} \label{sec_SPA}
	
\subsection{Static-Path Approximation} 
	
As previously stated, it is not possible to calculate the M1 strength function directly in the SMMC. Instead one can only calculate its Laplace transform, the imaginary-time response function $\langle \hat{\mathcal{O}}_{M1}(\tau) \hat{\mathcal{O}}_{M1}(0) \rangle$.   It is then necessary to carry out a numerical analytic continuation to find the strength function.  This can be accomplished using the maximum entropy method (MEM) but it requires a good choice for a prior strength function. Here we used the SPA as a prior distribution.
	
The SPA~\cite{Muhlschlegel1972, Alhassid1984, Lauritzen1988,Arve1988,Rossignoli:1992lpw,Alhassid:1992uyi,ATTIAS1997565} is an approximation in which the auxiliary fields are assumed to be time-independent, i.e.,  $\sigma_{\alpha}(\tau) = \sigma_{\alpha}$. This simplifies the functional integral in the HS transformation, eliminating the need to discretize Euclidean time. The SPA can therefore be interpreted as an approximation in which only static fluctuations around the mean field solution are included and quantum fluctuations are ignored. 

The M1 strength function can be calculated directly in the SPA without the need for analytic continuation using the adiabatic approximation~\cite{ATTIAS1997565,Rossignoli:1998yra,Mercenne2024}.
\begin{equation}
	S_{M1}(\omega;T) = \frac{\int d\sigma M(\sigma)Z_{\eta}(\sigma)S_{M1,\eta}(\omega;T,\sigma)}{\int d\sigma M(\sigma)Z_{\eta}(\sigma)},
\end{equation}
where $M(\sigma)$ is an integration measure, and $Z_{\eta}(\sigma)$ is the number-parity projected one-body partition function with $\eta=+1(-1)$ for even (odd) number parity~\cite{Fanto:2020wjr}.  The M1 strength for a given set of auxiliary fields $\sigma$ and number-parity $\eta$ is given by
\begin{equation}
	\label{eq_spa}
	\begin{aligned}
	&S_{M1,\eta}(\omega;T,\sigma) = S_{M1,\eta}^{(0)}(T,\sigma) \delta(\omega)	\\
	&+ \frac{1}{1-e^{-\beta \omega}} \sum_{\mu,k\ell} \frac{1}{2} |\hat{\mathcal{O}}^{\mu}_{M1,k\ell}|^2 (\tilde{f}^\eta_{\ell} - \tilde{f}^\eta_k) \delta (\omega - (\tilde{E}_k - \tilde{E}_{\ell})) \;,
	\end{aligned}
\end{equation}
where $\hat{\mathcal{O}}^{\mu}_{M1,k\ell} = \langle k | \hat{\mathcal{O}}^{\mu}_{M1} | \ell \rangle$ are the matrix elements of the $\mu$-component of the M1 operator, $\tilde{E}_k$ are the generalized quasiparticle energies, and $\tilde{f}^\eta_k$ are their associated number-parity projected thermal occupation numbers~\cite{Fanto:2020wjr}. The first term on the right side of Eq.~(\ref{eq_spa}) takes into account the zero-mode strength, which characterizes the transitions between degenerate states and is given by 
\begin{equation}
	S_{M1,\eta}^{(0)}(T,\sigma) = \frac{1}{2} \sum_{\mu} \sum_{kl,k'l'} \hat{\mathcal{O}}_{M1,kl}^{\mu*} \hat{\mathcal{O}}_{M1,k'l'}^{\mu} \langle \alpha_k^{\dagger} \alpha_l \alpha_{k'}^{\dagger} \alpha_{l'} \rangle_{\sigma,\eta}\;,
\end{equation}
where the sum is carried out only over pairs of degenerate states with $\tilde{E}_k = \tilde{E}_{\ell}$, and $\alpha_k,\alpha^\dagger_k$ are the quasi-particle annihilation and creation operators. In practice, rather than using the $\delta$-functions of Eq. (\ref{eq_spa}), Lorentzian functions of a finite width $\epsilon = 0.2$ MeV are used to produce smooth strength functions. This is consistent with previous SMMC calculations \cite{Fanto2024,Mercenne2024} and is similar to the energy bin widths used in conventional shell-model calculations \cite{Schwengner:2017cet,Karampagia:2017zgj}.     
	
An improvement to the SPA, called the SPA + RPA (random-phase approximation), accounts also for small-amplitude time-dependent quantum fluctuations of the auxiliary fields around each static configuation, improving the accuracy of the approximation at low temperatures; see Refs.~\cite{PhysRevC.42.R1830,PUDDU1991409,ATTIAS1997565,PhysRevLett.80.1853} for applications and further discussions of the SPA + RPA. The inclusion of the RPA requires the calculation of an  RPA correction factor for each static field configuration. Computation of this RPA correction factor require the construction and diagonalization of a large ($\mathcal{O}(N_s^2)$) matrix at each Metropolis update step which can make the computations significantly more costly. SPA+RPA calculations of the M1 strength function of heavy nuclei were carried out using a simplified pairing-plus-quadrupole Hamiltonian~\cite{Fanto2024} to reduce the computational cost. More numerically efficient methods based on contour integrals~\cite{PhysRevLett.83.280,PhysRevC.72.061306} and linear response theory~\cite{PhysRevLett.85.2260} have been developed and used in other settings, but have not yet been implemented for heavy nuclei.	
\subsection{Maximum-entropy method}

In order to numerically invert the SMMC imaginary-time response function data, we use the MEM~\cite{Jarrell:1996rrw}. The MEM determines the least biased (in the Bayesian probabilistic sense) M1 strength function $S_{M1}$ that reproduces the  SMMC response function $R_{M1}$ in Eq.~(\ref{eq_response}). 

More specifically, the MEM selects the strength function $S_{M1}$ that maximizes the objective function
\begin{equation}
	\label{eq_obj}
	\mathcal{Q}(S_{M1};\alpha) = \alpha \mathcal{S} - \frac{1}{2} \chi^2 \;,
\end{equation}
where $\chi^2$ is given by
\begin{equation}
	\chi^2 = (R^{\textrm{SMMC}}_{M1} - R_{M1})^T C^{-1} (R^{\textrm{SMMC}}_{M1} - R_{M1}) \;.
\end{equation}
In Eq. (\ref{eq_obj}) $C$ is the covariance matrix of the SMMC response function data and $\alpha$ is a free parameter that can in principle be chosen to maximize the probability function $\mathcal{P}[S_{M1}|R_{M1}^{\textrm{SMMC}}] \propto e^{\mathcal Q}$. The entropy functional $\mathcal{S}$ is (suppressing the arguments of the strength function)
\begin{equation}
	\mathcal{S} = -\int d\omega \left( S_{M1} - S_{M1}^{\textrm{prior}} - S_{M1} \ln (S_{M1}/S_{M1}^{\textrm{prior}}) \right) \;,
\end{equation}
where $S_{M1}^{\textrm{prior}}$ is a chosen prior strength function. 

The choice of a single value of $\alpha$ to maximize the probability function has been shown to produce good results when the distribution in $\alpha$ is sharply peaked. This is not always the case, so instead we apply Bryan's method~\cite{Bryan1990}, averaging over the selected strengths for all values of $\alpha$\begin{equation}
	S_{M1} = \int_0^{\infty} d\alpha S_{M1}^{\alpha}\mathcal{P}[\alpha|S_{M1},R_{M1}^{\textrm{SMMC}}],
\end{equation} 
where $S_{M1}^{\alpha}$ is the strength determined by maximizing the objective function (\ref{eq_obj}) for each given value of $\alpha$.

The MEM finds the strength function that best fits the response function data (i.e.,  minimizing $\chi^2$) but simultaneously keeps the strength close to the prior strength (i.e., maximizing ${\mathcal S}$).  Thus the accuracy of the MEM  depends on a good choice for the prior. Previous works~\cite{Fanto2024,Mercenne2024} have shown that for excitation energies near the neutron separation energy, the SPA provided a good approximation in the sense that the SPA response function is close to the calculated SMMC response function. At low temperatures, however other choices may be better. In particular, it was shown that while the SPA is still a good choice at low temperatures, the quasiparticle RPA provides a more accurate approximation for spherical nuclei near the ground state~\cite{Mercenne2024}.

We note that other methods based on Bayesian inference exist for carrying out the numerical analytic continuation besides the MEM; see, e.g., the method proposed in Ref.~\cite{Burnier:2013nla}.

The strength and response functions computed in the SMMC and SPA involve averaging over all configurations and include contributions from all spins possible in the model space. In order to study the spin-dependence of these functions, one could introduce a spin projection on the M1 transition operator to construct spin-dependent strength and response functions $S_{M1}(\omega;T,J)$ and $R_{M1}(\tau;T,J)$. This has not yet been implemented in the SMMC and SPA codes used here, necessitating the assumption of spin-independence mentioned in Sec.~\ref{sec_gamma}. 
 
\section{Results for odd-mass lanthanides} \label{sec_results}
	
\subsection{Uncertainty quantification and temperature dependence of the strength functions}

\begin{figure}[bth]
	\includegraphics[width=\linewidth]{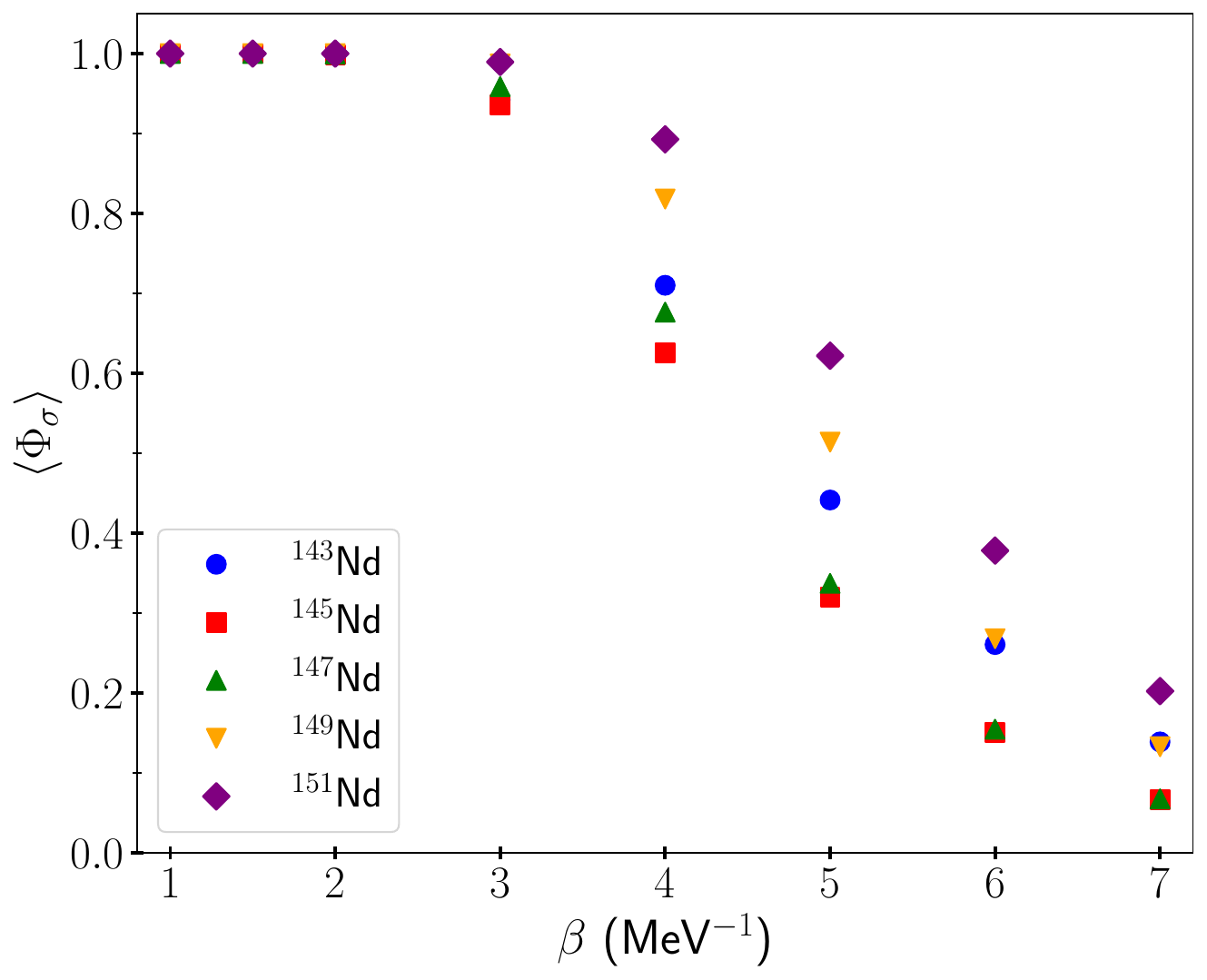}
	\caption{Average Monte-Carlo sign $\langle \Phi_{\sigma} \rangle$  as a function of the inverse temperature $\beta$ for the odd-mass neodymium isotopes using time slices of $\Delta \beta = 1/64$ MeV$^{-1}$.}
\label{fig_mc}
\end{figure} 

\begin{figure*}
	\includegraphics[width=\linewidth]{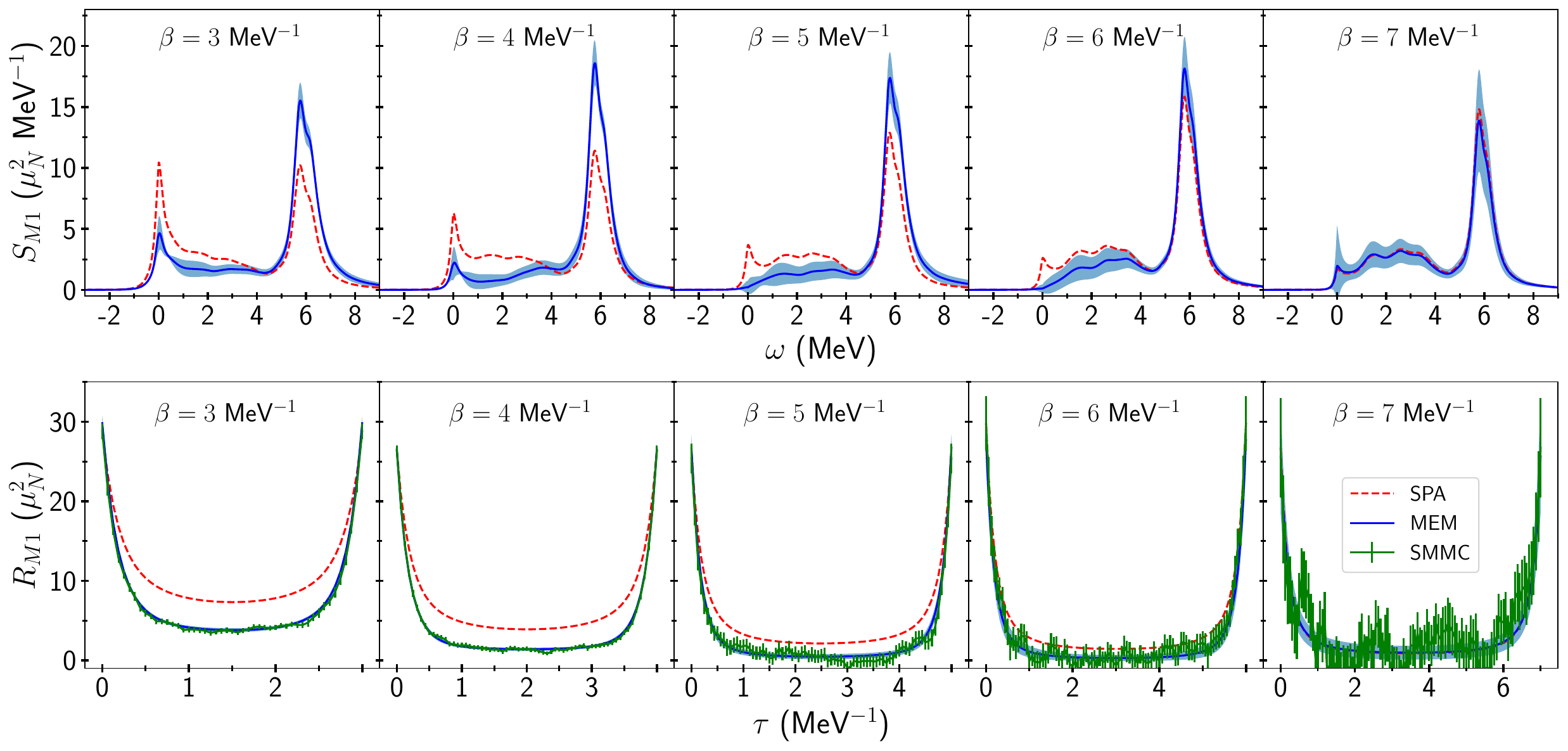}
	\caption{Finite-temperature M1 strength functions  $S_{M1}$ (top row) and response functions $R_{M1}$ (bottom row) of $^{145}$Nd for temperatures at which the Monte-Carlo sign of the SMMC is less than 1. The SPA prior strength and response functions are compared with the MEM results as well as the SMMC response function data. The SPA response functions are scaled to coincide with the SMMC response functions at $\tau=0$.  The blue bands denote the uncertainties in the strength functions.}
	\label{fig_SR}
\end{figure*}

We calculated the M1 $\gamma$SF for the odd-mass isotopes $^{143-151}$Nd and $^{147-153}$Sm at a range of inverse temperatures from values that correspond to the neutron separation energy of each nucleus $\beta \simeq 1.5$ MeV$^{-1}$ down to temperatures as low as $\beta \sim 6 - 7$ MeV$^{-1}$. Figure \ref{fig_mc} shows the average Monte-Carlo sign $\langle \Phi_\sigma \rangle$ as a function of $\beta$ for the odd-mass neodymium isotopes. We observe that the average sign begins to decrease rapidly below unity at $\beta \approx 2$ MeV$^{-1}$ for each of the odd-mass neodymium isotopes. Despite the rapid decrease of the average sign, the values remain non-zero within their uncertainty even at $\beta = 7$ MeV$^{-1}$.   Similar results are found for the odd-mass samarium isotopes. 

The odd-mass sign problem at low temperatures leads to large fluctuations in thermal observables such as the energy and it is a challenge to extract ground-state values. The uncertainties of the M1 response function also grow large at low temperatures (see, e.g., the rightmost panel of Fig.~\ref{fig_SR}).  It is important to quantify how these uncertainties affect the strength functions computed by the MEM and over what range of temperatures the results are reliable. 

For values of $\beta \le 3$ MeV$^{-1}$, we used  8,000 uncorrelated samples to calculate the response functions at each of the two values of $\Delta \beta$ ($1/32$ and $1/64$ MeV$^{-1}$), while for $\beta > 3$ MeV$^{-1}$ we used 32,000 uncorrelated samples. The SPA strength and response functions were computed using 8,000 samples for all values of $\beta$ since the SPA does not have a sign problem at low temperatures.
	
Direct error propagation of the uncertainties from the SMMC response function through the MEM would be difficult, and instead we carried out a jackknife analysis~\cite{Young2015}. In the jackknife method, the $i$-th jackknife average of an observable $O$ is computed from the average of all measurements excluding the $i$-th measurement $O_i$.  All $N$ jackknife averages are computed for a set of $N$ measurements, and the average and uncertainty of the set of jackknife averages are equivalent to the average value and uncertainty that would be computed via direct error propagation. Here we computed the jackknife averages of the SMMC response functions and apply the MEM to each of them. The average and uncertainty of these sets of M1 $\gamma$SFs and response functions were then computed and taken as the central value and error of the calculation. Performing the MEM for a large number of response functions  is computationally expensive, and individual measurements are instead averaged into $N_{\textrm{block}}$ blocks and only $N_{\textrm{block}}$  jackknife averages are used. For the results presented in this work, we used $N_{\textrm{block}} = 200$. 

We demonstrate the jackknife method in Fig.~\ref{fig_SR} for $^{145}$Nd for the M1 strength function (top row) and response function (bottom row) at low temperatures for which the uncertainties can become large due to the odd particle-number sign problem. We observe that the statistical fluctuations of the SMMC response function grows rapidly with increasing $\beta$, particularly at $\beta \ge 5$ MeV$^{-1}$, where the average Monte-Carlo sign is less than 0.4. This uncertainty in the SMMC response function manifests in both the MEM strength and response via the $\chi^2$ term in the objective function (\ref{eq_obj}).

It is interesting to note that for $\beta = 7$ MeV$^{-1}$, where the uncertainties in the MEM $\gamma$SF results do not seem to grow compared with the $\beta = 6$ MeV$^{-1}$ results despite the much noisier SMMC response function data. When the uncertainty is the SMMC response function becomes sufficiently large, the value of $\chi^2$ shrinks, giving it little weight in the objective function. As a result, the statistical fluctuations in the SMMC data do not propagate into the MEM results. In this case, the MEM effectively becomes insensitive to the SMMC data and is entirely biased towards the SPA prior strength.  Indeed for $\beta = 7$ MeV$^{-1}$, Fig.~\ref{fig_SR} shows that the SPA and MEM strength functions are nearly identical. 

The temperatures of most interest to us are those at which the average excitation energy $\langle E_x \rangle$ is near the neutron separation energy $S_n$. This is the initial excitation energy of the compound nucleus formed in the Oslo-method experiments. The odd-mass rare-earth nuclei discussed here have $S_n \sim 5 - 6$ MeV, corresponding to $\beta \sim 1.5$ MeV$^{-1}$. These temperatures are sufficiently high so that the sign problem does not yet manifest. 

\begin{figure*}[bth]
	\includegraphics[width=\linewidth]{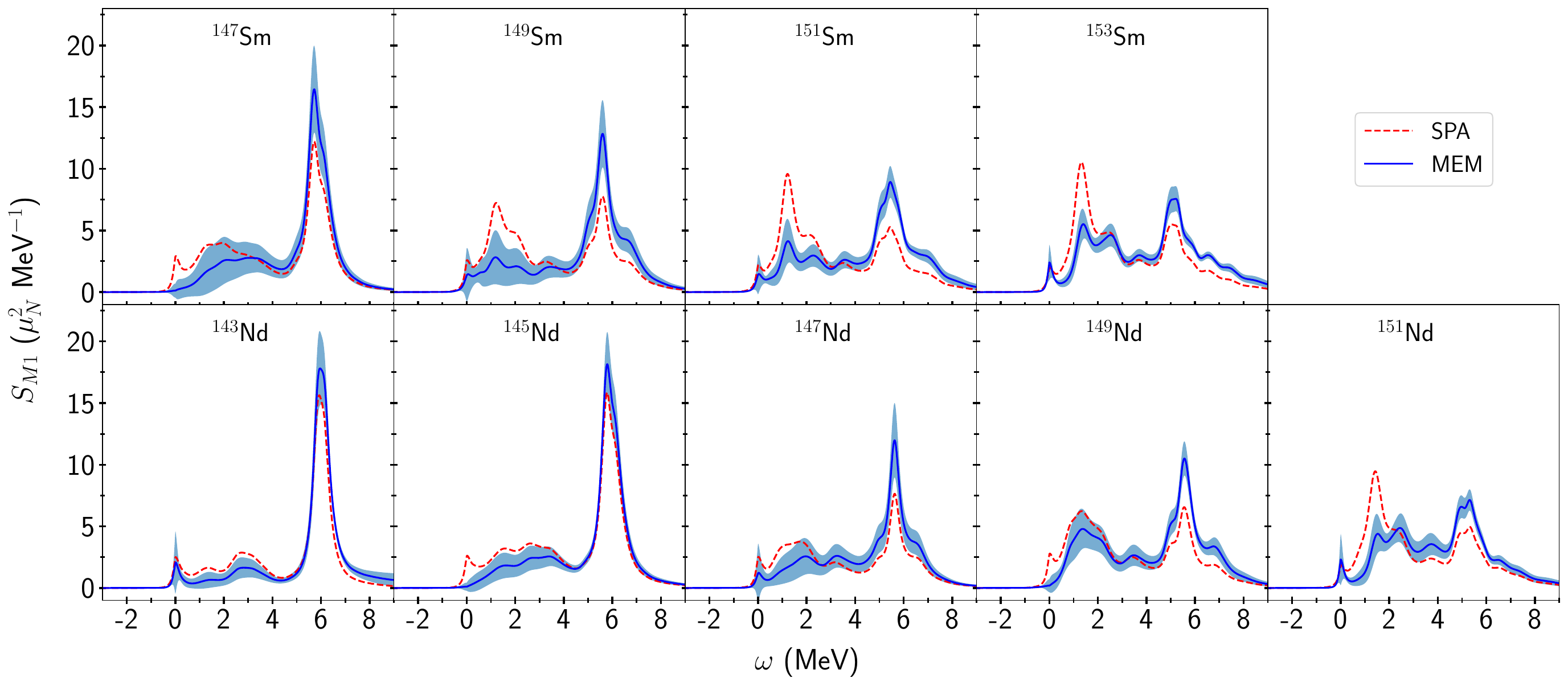}
	\caption{M1 strength functions $S_{M1}$  near the ground state ($\beta = 6 $ MeV$^{-1}$) for the odd-mass $^{\textrm{147-153}}$Sm (top row) and $^{\textrm{143-151}}$Nd (bottom row) isotopes. The SPA prior strengths  (dashed red lines) are compared with the MEM  strengths (solid blue lines with bands around them to mark the statistical errors).}
	\label{fig_SF}
\end{figure*}

\begin{figure*}[bth]
	\includegraphics[width=\linewidth]{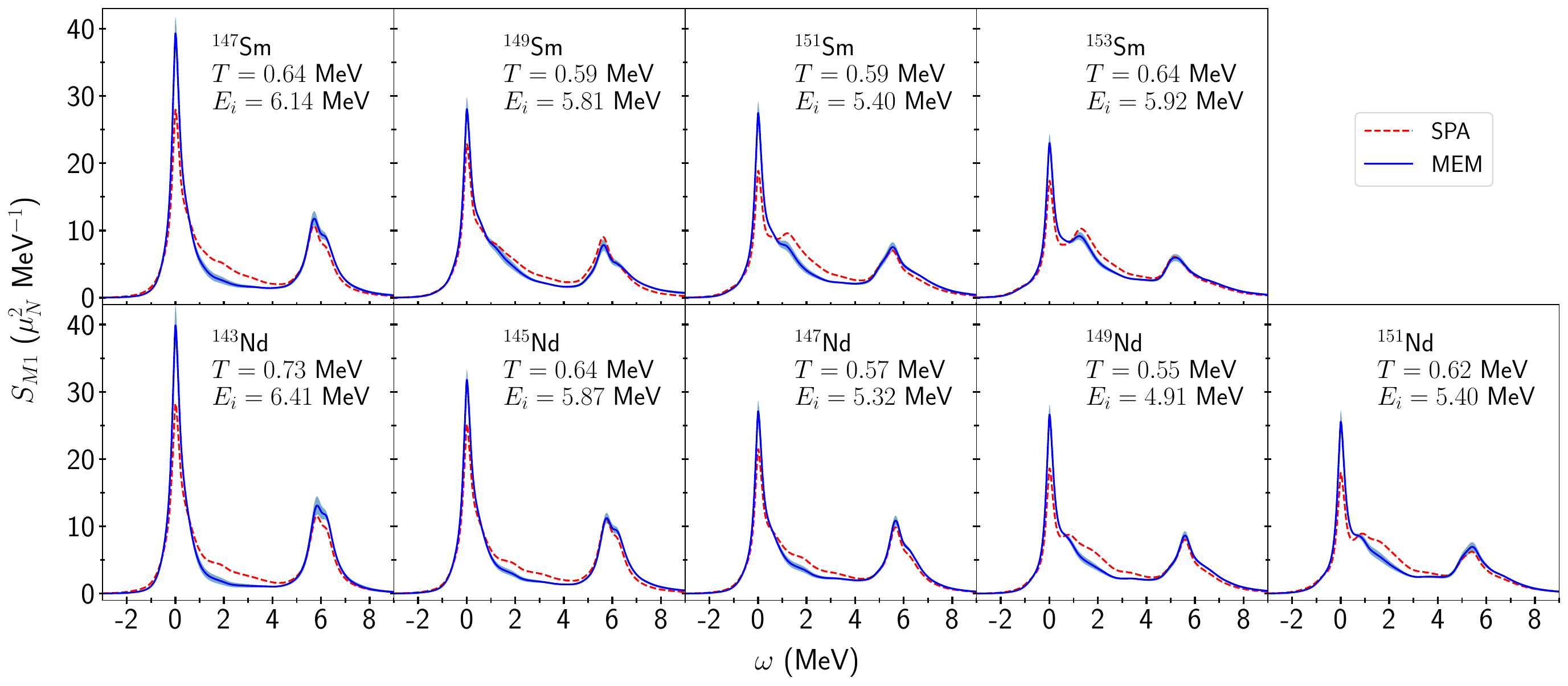}
	\caption{Same as in Fig.~\ref{fig_SF}, but at temperatures that correspond to the neutron separation energy.}
	\label{fig_SF2}
\end{figure*}

\subsection{M1 strength functions $S_{M1}$}

In this section we present results for the M1 strength functions in odd-mass samarium and neodymium isotopes at low temperatures (near the ground state) and at temperatures that correspond to the neutron separation energy. 

\subsubsection{M1 strength functions near the ground state}

In Fig.~\ref{fig_SF} we show the M1 strength functions  $S_{M1}$  at a low temperature $\beta= 6$ MeV$^{-1}$ for which the statistical errors are still manageable for the odd-mass  $^{\textrm{147-153}}$Sm (top row) and $^{\textrm{143-151}}$Nd (bottom row) isotopes. The SPA strengths are shown by the dashed red lines, while the MEM M1 $\gamma$SFs are shown by the solid blue lines with the statistical error forming a band around their average.

We observe a multi-peaked structure around $\omega \approx 1- 4$ MeV that becomes more pronounced in the crossover from spherical to deformed isotopes. We interpret this structure as the scissors resonance in which the proton and neutron clouds oscillate against each other like scissors blades.  

We further interpret the pronounced peak seen near $\omega \approx 6$ MeV as the M1 spin-flip mode. This mode describes the flip of one or more of the nuclear spins. It becomes more fragmented and its peak height reduces as we add more neutrons in the crossover from spherical to deformed isotopes~\cite{Mustonen:2018ody}. 

\subsubsection{M1 strength functions near the neutron separation energy}

In Fig.~\ref{fig_SF2} we show the $S_{M1}$ strength functions at temperatures that correspond to the neutron separation energy for the same odd-mass isotopes as in Fig.~\ref{fig_SF}, comparing the SPA strengths (dashed red lines) with the MEM strengths (solid blue lines).  We observe the appearance of a large peak centered at $\omega=0$ that is not present near the ground state strength.   We will identify this $\omega=0$ peak as the LEE (see Sec.~\ref{sec_lee}).  

As neutron number increases, the amplitude of the LEE peak decreases and a smaller secondary peak appears in the deformed isotopes (see in particular for $^{\textrm{151-153}}$Sm $^{\textrm{149-151}}$Nd). This secondary peak is interpreted as the scissors mode built on top of excited states.  We also observe a spin-flip mode at $\omega \sim 6$ MeV that is built on top of excited states. This mode at finite temperature exhibits less structure than its counterpart near the ground state.

\subsection{The deexcitation $\gamma$SF $f_{M1}$} \label{sec_lee}
	
The strength function  $S_{M1}$ can be converted to the deexcitation $\gamma$SF $f_{M1}$ using Eq.~(\ref{eq_gammaSF}). In the experiment a range of spins is populated in the reaction that produces the relevant compound nucleus. In the following we ignore the spin dependence of $S_{M1}$ and use relation (\ref{eq_gammaSF}) to convert $S_{M1}(T;\omega=-E_\gamma)$ to $f_{M1}(E_i; E_\gamma)$.  

\subsubsection{Level densities of odd-mass isotopes}

The conversion from $S_{M1}$ to $f_{M1}$ requires knowledge of the level density $\tilde\rho(E_x)$ of the appropriate compound nucleus. In contrast to the state density, the level density  does not include the $2J+1$ magnetic degeneracy of each level with spin $J$.  The level density is a function of the excitation energy $E_x=E-E_0$ and the determination of this excitation energy requires accurate knowledge of the ground-state energy $E_0$. 

The projection onto an odd number of fermions introduces a sign problem in SMMC at low temperatures.  Because of this odd particle-number sign problem, SMMC calculations of the odd-mass lanthanides have been limited up to $\beta \sim 4-5$ MeV$^{-1}$, while calculations for their even-mass neighbors have been carried out to at least $\beta = 20$ MeV$^{-1}$~\cite{Ozen:2012gq}. 
 Here, we used the values of the ground-state energies of the odd-mass isotopes $^{\textrm{149-153}}$Sm and $^{\textrm{143-151}}$Nd  extracted in Ref.~\cite{Ozen:2013xza} using a method that is based experimental data. We used a the same method to extract a ground-state energy of  $E_0 = -224.53$ MeV for$^{\textrm{147}}$Sm. 
 
We note that other methods were developed for calculating the ground-state energy of odd-mass nuclei that do not rely on experimental data. One method,  based on imaginary-time Green's function of neighboring even-mass nuclei, was used successfully in mid-mass nuclei~\cite{Mukherjee:2012yn}, but becomes too time consuming in heavy nuclei. Another method for heavy nuclei  that does not rely on experimental data, the partition function extrapolation method, was developed in Ref.~\cite{Alhassid2023} and gives values for $E_0$ that are in agreement with the values calculated by the method of Ref.~\cite{Ozen:2013xza}. 

\begin{figure}[bth]
\includegraphics[width=\linewidth]{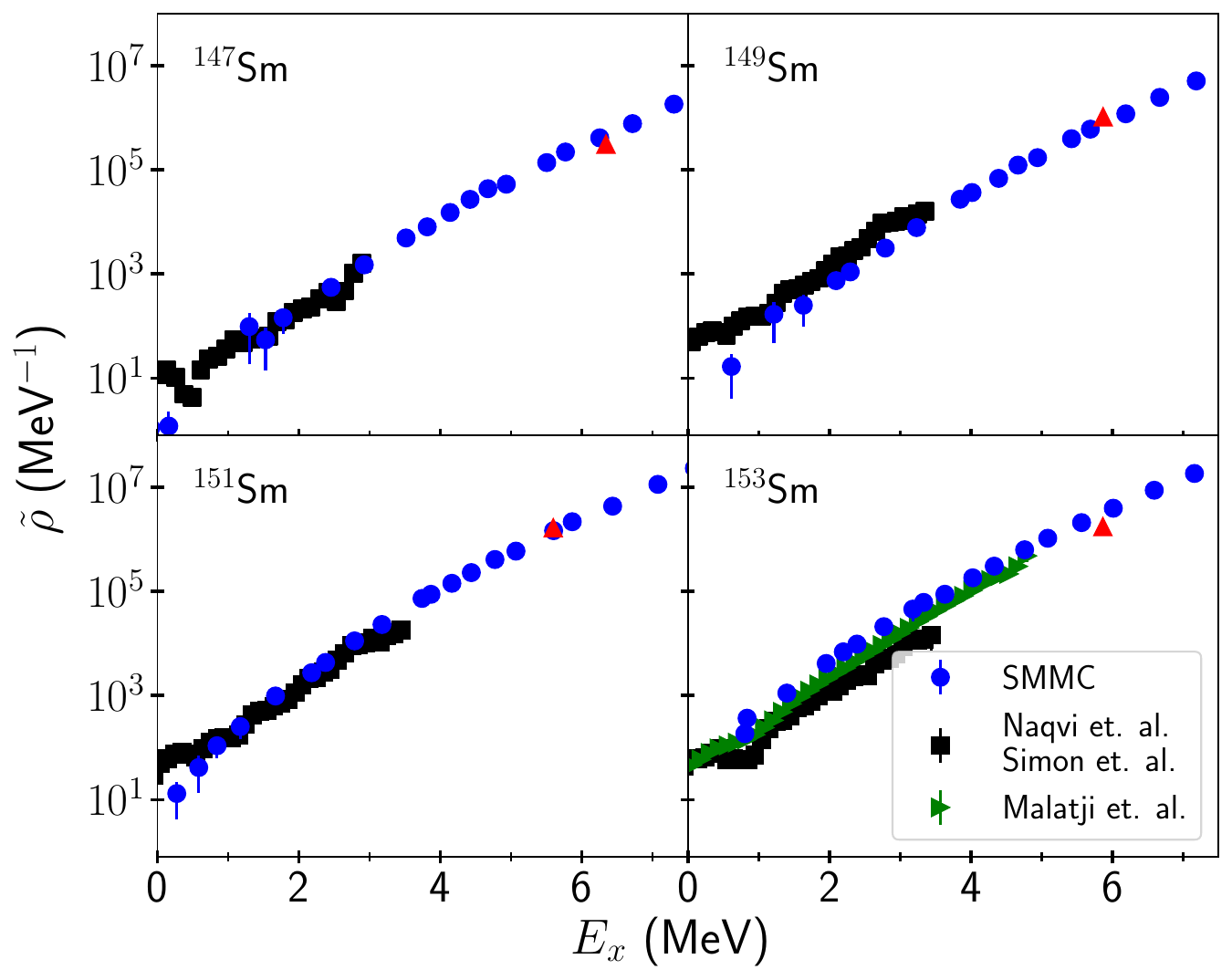}
\caption{Level densities $\tilde\rho(E_x)$ for the odd-mass samarium isotopes $^{147-153}$Sm. The SMMC level densities (blue circles with statistical errors) are compared with the experimental results of Refs.~\cite{Simon:2016mif,Naqvi:2019odv} (black squares) and Ref.~\cite{Malatji:2021nar} (green triangles). The red triangles are the level densities at the neuron separation energy determined from neutron resonance data.}
\label{fig_rho}
\end{figure} 

In order to compute the level density $\tilde \rho$, it is necessary to count the contribution from each level with spin $J$ just once without the $2J+1$ magnetic degeneracy. For each level in an odd-mass nuclei, each state with spin component $M=1/2$  appears only once, thus projecting onto states with $M=1/2$ amounts to a removal of the magnetic degeneracy factor, i.e., $\tilde \rho =\rho_{M=1/2}$~\cite{Alhassid2013}. This is accomplished using the spin projection method introduced in Ref.~\cite{Alhassid:2006np}.

In Fig.~\ref{fig_rho} we show the SMMC level densities of the samarium isotopes in comparison with those determined from experiments~\cite{Naqvi:2019odv,Malatji:2021nar}. For $^{\textrm{147-151}}$Sm, we find that  the SMMC reproduces well the experimental level densities. The SMMC level density for $^{\textrm{153}}$Sm is in good agreement with the experimental results of Ref.~\cite{Malatji:2021nar} while it is higher by a roughly constant factor than the experimental level density of Ref.~\cite{Naqvi:2019odv}. 

For the odd-mass neodymium isotopes, we use the level densities calculated in Ref.~\cite{Guttormsen:2020nnp}.

\subsubsection{Dependence of the M1 $\gamma$SF on initial energy}

\begin{figure}[b]
\includegraphics[width=\linewidth]{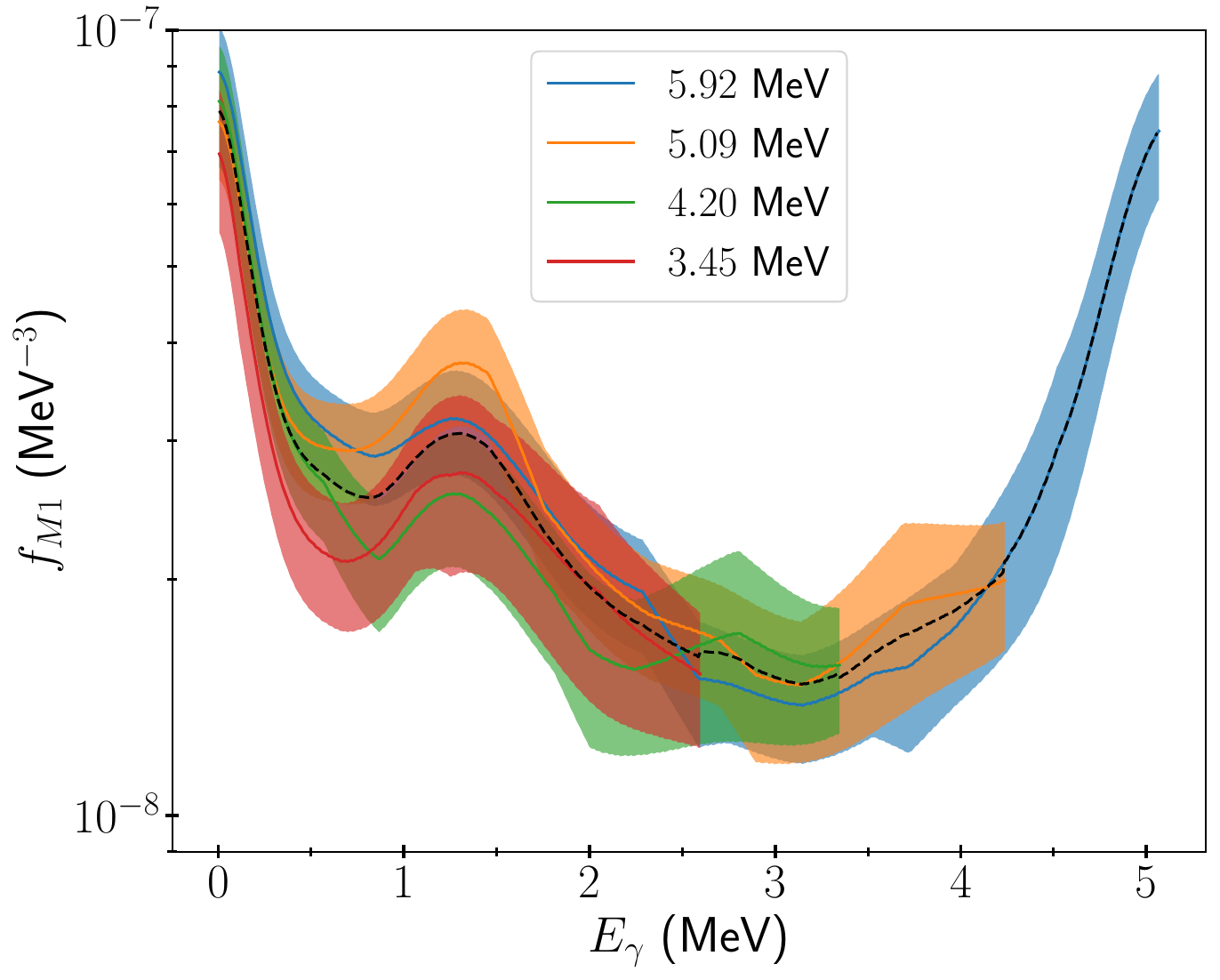}
\caption{Deexcitation M1 $\gamma$SFs in $^{153}$Sm computed at different initial excitation energies $E_i$ shown in the legend. The dashed black line is the average of all these curves.}
\label{fig_ei}
\end{figure}

In the Oslo experiments, the M1 $\gamma$SF is typically an average over an initial energy window $E_{min} < E_i < S_n$, where $E_{min}$ is a lower bound $\sim 2-3$ MeV and $S_n$ is the neutron separation energy.  Figure \ref{fig_ei} shows the M1 $\gamma$SF computed at several values of the initial energy at and below the neutron separation energy in $^{153}$Sm. Our  results confirm that the M1 $\gamma$SF depends only weakly on the initial energy $E_i$ in the energy range of interest. In particular, features such as the slope of the LEE and the center and width of the SR peak show no systematic variation with $E_i$. In fact, the entire M1 $\gamma$SF is essentially independent of $E_i$ up to the uncertainties.

\begin{figure*}
	\includegraphics[width=\linewidth]{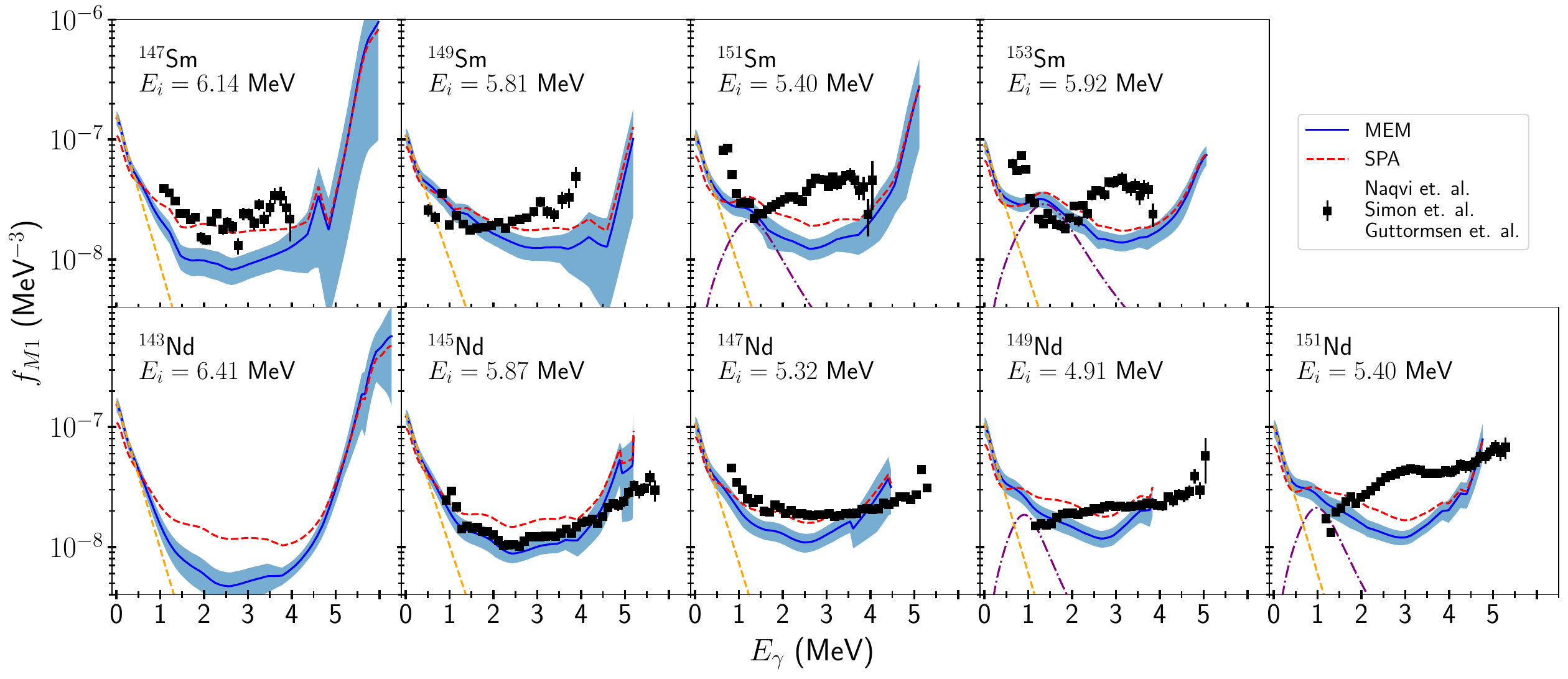}
	\caption{Deexcitation M1 $\gamma$SFs for the odd-mass samarium (top row) and neodymium (bottom row) isotopes computed at average initial energies $E_i$ that correspond to their neutron separation energies. The MEM results (solid blue lines with bands indicating the uncertainties) are compared with the SPA strengths (dashed red lines).  The black squares are the experimental data~\cite{Simon:2016mif,Naqvi:2019odv,Guttormsen:2022ccz} for the total (both E1 and M1) $\gamma$SFs when available. The dashed orange lines are the fit of the MEM $f_{M1}$ to the form (\ref{eq_lee}) of the LEE, and the  dash-dotted purple lines are the fits of the MEM results to the form  (\ref{eq_sm}) of the SR after subtracting the LEE fits.}
\label{fig_gamma}
\end{figure*}

\begin{table*}
	\caption{Comparison of the LEE and SR fit parameters calculated within the SMMC + MEM and those measured in experiments. The experimental results are taken from Ref.~\cite{Guttormsen:2022ccz} for the neodymium isotopes and from Refs.~\cite{Simon:2016mif,Naqvi:2019odv} for the samarium isotopes.}
	\begin{tabular}{c|ccccc|ccccc}
		\hline
		\hline
		Nucleus & \multicolumn{5}{c|}{SMMC + MEM} & \multicolumn{5}{c}{Experiment} \\
		& $C_0$ & $\kappa$ & $\sigma$ & $\omega$ & $\Gamma$  &  $C_0$  & $\kappa$ & $\sigma$ & $\omega$ & $\Gamma$   \\
		& ($10^{-7}$ MeV$^{-3}$) & (MeV$^{-1}$) & (mb) & (MeV) & (MeV) & ($10^{-7}$ MeV$^{-3}$) & (MeV$^{-1}$) & (mb) & (MeV) & (MeV) \\
		\hline
		$^{143}$Nd & 1.64(7) & 2.83(18) &&&& &  & &  & \\
		$^{145}$Nd & 1.29(6) & 2.53(19) &&&& 1.26(21) &1.9(2)\footnote[1]{Experimental value of the slope parameter $\kappa$ is the average value for $^{\textrm{142,144-147}}$Nd.} & & & \\
		$^{147}$Nd & 1.09(6) & 2.69(23) &&&& 1.65(11) &1.9(2)\footnotemark[1] & 0.12(4) & 2.22(16)& 1.0(5)  \\
		$^{149}$Nd & 1.08(6) & 2.88(24) & 0.21(3) & 1.05(15) & 1.09(49) &  &  & 0.12(4) & 2.37(16) & 1.9(5) \\
		$^{151}$Nd & 1.04(6) & 2.85(23) & 0.26(3) & 1.14(14) & 1.20(45) & &  & 0.64(27) & 2.95(25) & 1.1(6) \\
		$^{147}$Sm & 1.60(7) & 2.89(18) &&&& 10(5) &3.2(10)  & &  &  \\
		$^{149}$Sm & 1.11(5) & 2.42(20) &&&& 20(10) &5.0(10)  &  & &   \\
		$^{151}$Sm & 1.11(6) & 2.57(20) & 0.33(2) & 1.43(7) & 1.45(24) & 20(10) &5.0(5)   & 0.6(2) & 3.0(2) & 1.1(3)  \\
		$^{153}$Sm & 0.93(4) & 2.55(17) & 0.49(2) & 1.57(6) & 1.60(16) & 20(10)\footnote[2]{The LEE was observed experimentally in $^{\textrm{153}}$Sm in Ref. \cite{Simon:2016mif}, but not in Ref. \cite{Malatji:2021nar}.} &5.0(10)\footnotemark[2]  & 0.6(2) & 3.0(2)& 1.1(2)  \\
		\hline
		\hline
	\end{tabular}
	\label{tab1}
\end{table*}

\subsubsection{LEE and scissors resonance}
		
We converted the strength functions $S_{M1}$ for the odd-mass samarium and neodymium isotopes at the neutron separation energy (shown in Fig.~\ref{fig_SF2}) to $\gamma$SF $f_{M1}$ using Eq.~(\ref{eq_gammaSF}) and the SMMC level densities.  In Fig.~\ref{fig_gamma} we compare these M1 $\gamma$SF to the $\gamma$SF extracted in Oslo experiments (black squares). The SPA results are shown by the dashed red lines and the MEM results (with the SPA as prior) are shown by the solid blue lines with the blue bands indicating the statistical uncertainties. These bands are relatively narrow at low $E_{\gamma}$ since the sign problem at the temperatures that correspond to the neutron separation energies is moderate, but are wider at large $E_{\gamma}$ due to the uncertainties in the level densities at low excitation energies. We note that the experimental $\gamma$SFs include contributions from both M1 and E1 and a more detailed comparison between theory and experiment will require the calculations of the E1 $\gamma$SF. 

We clearly observe in the calculated $\gamma$SFs the 'upbend' structure of the LEE in all of the isotopes at $\gamma$-ray energies below $\sim 2$ MeV. The LEE peaks originate in the $\omega=0$ peaks seen in the $S_{M1}$ strength functions (see Fig.~\ref{fig_SF2}).  The LEE structure is the feature in the M1 $\gamma$SF that is the least sensitive to the choice of prior strength when applying the MEM. The calculated LEE structures are in overall agreement with the experiment. The exceptions are $^{\textrm{151,153}}$Nd, for which no LEE is observed experimentally. However, this may be due to the experimental limitation to access $\gamma$-ray energies  $E_{\gamma}$ below $\sim 1$ MeV.

It was proposed that the LEE can be well described by an exponential form~\cite{Schwengner:2013ora}
\begin{equation}
	f_{M1}^{\textrm{LEE}}(E_{\gamma}) = C_0 e^{-\kappa E_{\gamma}} \;.
	\label{eq_lee}
\end{equation}
We find that this exponential form indeed provides a good fit to the $\gamma$SF $f_{M1}$ of the odd-mass samarium and neodymium isotopes for $E_{\gamma} \lesssim 1$ MeV. 

In the more deformed nuclei, a smaller, secondary peak can be seen at $E_{\gamma}$ just below 2 MeV. We interpret this peak as the scissors resonance (SR) at finite temperature.  
To isolate the SR, we subtract from the MEM $f_{M1}$ the fitted form (\ref{eq_lee}) of the LEE  and fit the residual strength to a standard Lorentzian~\cite{CAPOTE20093107}
\begin{equation}
	f_{M1}^{\textrm{SR}}(E_{\gamma}) = \frac{1}{3 \pi^2 (\hbar c)^2} \frac{\sigma E_{\gamma} \Gamma^2}{(E_{\gamma}^2 - \omega^2)^2 + E_{\gamma}^2 \Gamma^2} \;,
	\label{eq_sm}
\end{equation}
where $\sigma$, $\omega$, and $\Gamma$ are, respectively, the strength, center, and width of the SR.  We perform this SR fit only for $^{\textrm{149,151}}$Nd and $^{\textrm{151,153}}$Sm where the SR is clearly visible within the M1 $\gamma$SF. 

Table \ref{tab1} lists the fitted values of the LEE and SR parameters and compares them to those fitted to experimental data. Our theoretical results indicate that the slope parameter $\kappa$ of the LEE is roughly independent of the nucleus. While the experimental results show this to be the case within the samarium isotopes, they show a significant difference between the samarium and neodymium isotope chains.  The experimental heights $C_0$ of the LEE peaks at $E_\gamma \rightarrow 0$  in $^{145,147}$Nd are in overall agreement with their theoretical values. The theoretical values of $C_0$ within the samarium chain are similar to their values within the neodymium chain, but the experimental values of $C_0$ for the samarium isotopes are significantly larger than their values for the neodymium isotopes.  We find the LEE slope parameter $\kappa$ be approximately constant across both isotopic chains.

 The SR strengths $\sigma$ and widths $\Gamma$ show reasonable agreement between theory and experiment given the large uncertainties in the experimental values. The largest discrepancy is seen in the SR centers $\omega$ where the theoretical values are lower than the values extracted from the experiments. 

This discrepancy in the SR central energy may be an effect of the truncated model space. Previous work studying nuclear deformation in the lanthanides using the same model space and Hamiltonian~\cite{Mustonen:2018ody} found that a phenomenological factor must be included to account for core polarization effects and generate sufficient deformation.  Because the SR is the counter-rotation of the deformed proton and neutron clouds, the lack of sufficiently large deformation may reduce the SR energy. 

\begin{figure}[t]
	\includegraphics[width=\linewidth]{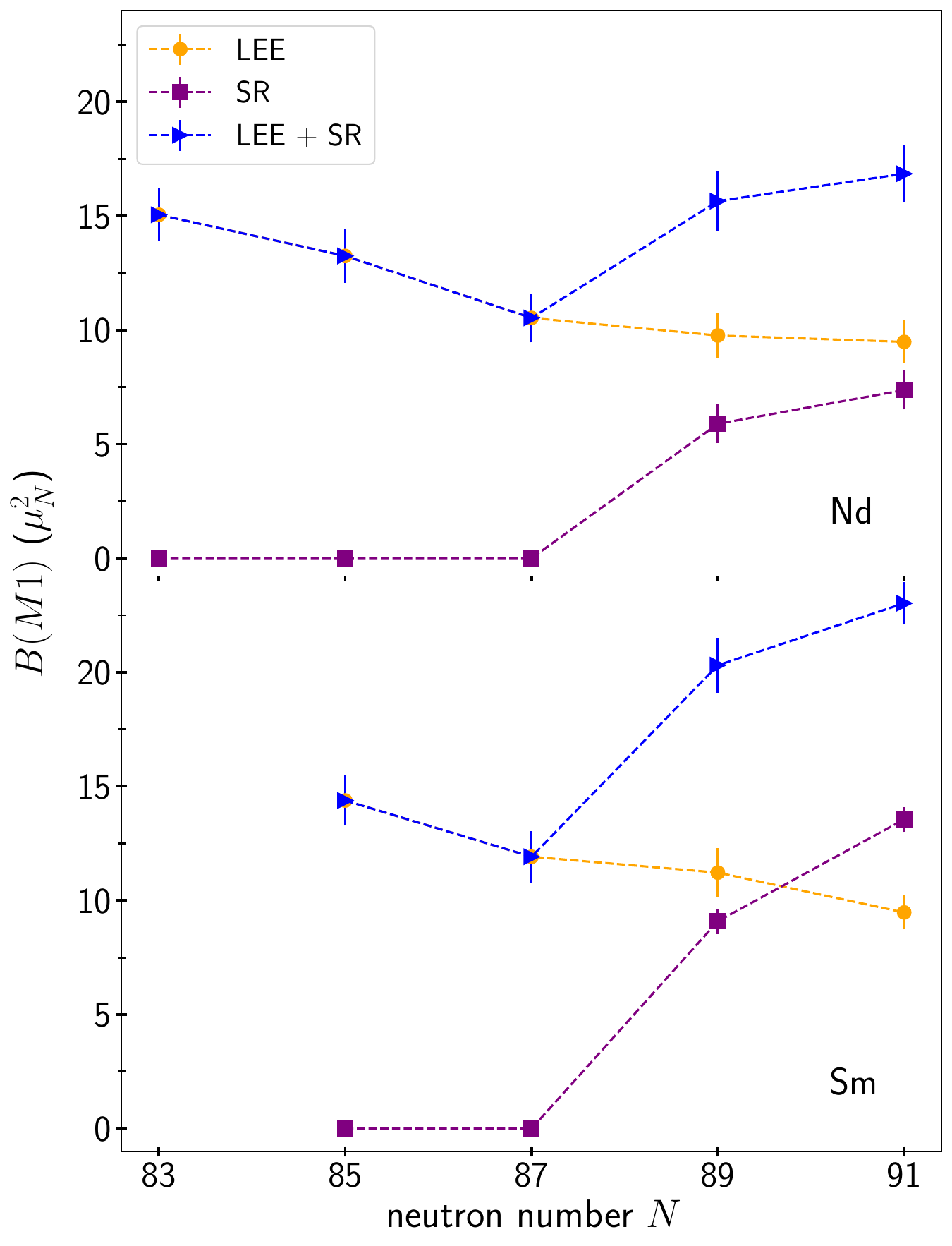}
	\caption{Integrated $B(M1)$ contribution (\ref{eq_bm1}) from the LEE and SR as well as their sum as a function of the neutron number $N$ for the isotopic chains of neodymium and samarium nuclei.}
	\label{fig_bm1}
\end{figure}

In order to compare the total M1 strengths of the LEE and SR across each isotopic chain, we calculated the total integrated $B(M1)$ strengths for each of these structures using
\begin{equation}
\label{eq_bm1}
B(M1) = \frac{27(\hbar c)^3}{16 \pi} \int_0^{S_n} f_{M1}(E_{\gamma}) dE_{\gamma} \;,
\end{equation}
where $f_{M1}$ is the fitted form of both the LEE (\ref{eq_lee}) and the SR (\ref{eq_sm}). Across each isotopic chain, the integrated strength of the LEE decreases as the neutron number increases, while the SR emerges in the deformed nuclei and its strength increases with neutron number. These integrated $B(M1)$ strengths and their sums are shown in Fig.~\ref{fig_bm1} for the neodymium chain (top panel) and the samarium chain (bottom panel). The  decrease in the LEE strength with neutron number is compensated for by the increase in the SR strength, leading to a total strength that varies weakly within each chain except for a dip at $N=87$. We note that in these calculations, there is some additional systematic uncertainty associated with the choice of the range of $E_{\gamma}$ over which the fits are performed. 

Conventional CI shell-model calculations in isotopic chains of mid-mass nuclei found the total integrated strengths are nearly independent of neutron number~\cite{Schwengner:2017cet,Frauendorf:2021gfd}. The experimental findings in the neodymium isotope chain do not support this claim~\cite{Guttormsen:2022ccz}. 

An important consideration when comparing the theoretical and experimental results for the $\gamma$SF is that the experimental results are for the total $\gamma$SF, containing contributions from both E1 and M1 transitions. While it is well established that the scissor and spin-flip modes are magnetic dipole in nature, the LEE has only been determined experimentally to be of a dipole nature. Although this work and previous CI shell-model calculations show an M1 contribution to the LEE, other theoretical results have suggested that thermal excitations in the continuum may contribute to an enhancement of low-energy E1 transitions~\cite{Litvinova:2013oda}.  

Another consideration is the difference between the theoretical calculations done at fixed temperatures (canonical ensemble) and the experiment performed at fixed initial energy (microcanonical ensemble). Previous calculations of  the M1 $\gamma$SF in neighboring even-mass nuclei~\cite{Fanto2024,Mercenne2024}  found that the LEE slope parameter $\kappa$ is roughly constant over a range of initial average energies $E_i$ above a low value of $\sim 2$ MeV. This independence is also seen in conventional CI shell-model calculations in lighter nuclei~\cite{Karampagia:2017zgj}. The weak dependence of the entire M1 $\gamma$SF with respect to the initial energy $E_i$, seen in Fig.~\ref{fig_ei}, justifies the approximation of replacing the microcanonical ensemble relevant to experiments by a canonical ensemble.
		
\section{Summary and Discussion} \label{sec_sum}
	
In this work, we calculated the M1 $\gamma$SFs for the odd-mass isotopes $^{\textrm{143-151}}$Nd and $^{\textrm{147-153}}$Sm up to their neutron separation energies. We used the MEM to carry out a numerical analytic continuation of the SMMC imaginary-time M1 response function with the SPA M1 strength function as prior to obtain a reliable M1 strength function. This work is the first such calculations in heavy odd-mass nuclei and shows that the $\gamma$SF can be accurately calculated despite the presence of a Monte Carlo sign problem at low temperatures that originates in the projection onto an odd number of neutrons. The LEE and SR were found to vary with deformation as neutron number increases along an isotopic chain. We identify a LEE in the $f_{M1}$ $\gamma$SF of all of the isotopes studied here but with decreasing strength as the neutron number is increased, while the SR is visible only in the more neutron-rich deformed isotopes.

Further improvements in the calculations of the M1 $\gamma$SF in the SMMC framework can be made by including spin projection to determine the spin dependence of the $M1$ strength function. 
 	
\begin{acknowledgments}
	
We thank P. Fanto for the use of the SPA and MEM codes he developed and A. C. Larsen for helpful discussions of the experimental results. This work was supported in part by the U.S. DOE grant No.~DE-SC0019521. The calculations used resources of the National Energy Research Scientific Computing Center (NERSC), a U.S. Department of Energy Office of Science User Facility operated under Contract No.~DE-AC02-05CH11231. We thank the Yale Center for Research Computing for guidance and use of the research computing infrastructure.

\end{acknowledgments}
	

%
	
\end{document}